\documentclass[aps,prd,superscriptaddress,showpacs,nofootinbib,floatfix,twocolumn]{revtex4}
\usepackage{graphicx} 
\usepackage{amsmath}
\newcommand{\s}[1]{{\ooalign{\hfil/\hfil\crcr$#1$}}}
\newcommand{\beq}{\begin{equation}}
\newcommand{\eeq}{\end{equation}}
\newcommand{\beqa}{\begin{eqnarray}}
\newcommand{\eeqa}{\end{eqnarray}}
\newcommand{\nnb}{\nonumber}

\begin{document}

%\preprint{}

\title{Probing vacuum birefringence
under a high-intensity laser field with gamma-ray polarimetry at the GeV scale}


\author{Yoshihide Nakamiya}
\affiliation{Advanced Research Center for Beam Science, Institute for Chemical Research, Kyoto University, Gokasho, Uji, Kyoto 611-0011, Japan}
\author{Kensuke Homma${}^*$}
\affiliation{Graduate School of Science, Hiroshima University, Kagamiyama, Higashi-Hiroshima, Hiroshima 739-8526, Japan} 
\email[corresponding author: ]{khomma@hiroshima-u.ac.jp}
\affiliation{International Center for Zetta-Exawatt Science and Technology, Ecole Polytechnique, Route de Saclay, F-91128 Palaiseau Cedex, France}
%\thanks{}
%\affiliation{}


\date{\today}

\begin{abstract}
Probing vacuum structures deformed by high intense
fields is of great interest in general.
In the context of quantum electrodynamics (QED), the vacuum exposed by 
a linearly polarized high-intensity laser field is expected 
to show birefringence.
We consider the combination of a 10~PW laser system to pump
the vacuum and 1~GeV photons to probe the birefringent effect.
The vacuum birefringence can be measured via the polarization flip of 
the probe $\gamma$-rays which can also be interpreted as phase retardation
of probe photons.
We provide theoretically how to extract phase retardation of GeV probe photons
via pair-wise topology of the Bethe-Heitler process in a polarimeter
and then evaluate the measurability of the vacuum birefringence 
via phase retardation given a concrete polarimeter design 
with a realistic set of laser parameters 
and achievable pulse statistics.
\end{abstract}

% 12.20.-m      Quantum electrodynamics
% 12.20.Fv      Experimental tests (for optical tests in quantum electrodynamics, see 42.50.Xa)
%41.75.Jv Laser-driven acceleration
% 25.75.Cj	Photon, lepton, and heavy quark production in relativistic heavy ion collisions
% 29.30.Kv	X- and gamma-ray spectroscopy
%42.62.-b Laser applications.
%\keywords{}

\maketitle

\section{Introduction}
The quantum nature of the vacuum in various extreme conditions is 
an intriguing subject to explore. 
The vacuum structure can be deformed by the existence of external fields,
such as gravitational fields~\cite{Drummond} and 
electromagnetic fields~\cite{Toll}.
It can also be modified by special boundary conditions 
via Casimir effects~\cite{Scharnhorst}. 
Common observables in these vacuum states are 
the dispersion relation for probe photons and 
polarization dependence~\cite{Shore}.
One interesting question is how much these properties 
differ between the present and the early Universe
when field densities were extremely high 
for certain boundary conditions~\cite{VSL}.
These properties are governed by the virtual quanta contained in the vacuum 
immersed in these intense fields, and 
the dominant virtual quanta differ depending on the dynamics. Therefore,
the energy scale of the probe photon is an important factor as well as the external field strength.

Understanding the interactions of probe photon with external fields requires 
non-trivial field theoretical treatments in the non-perturbative regime, 
where summing up all-order Feynman diagrams is necessary. 
Among various types of intense fields, the theoretical predictions 
in the simplest QED case naturally become the first candidates
to be thoroughly tested by laboratory experiments. 
Although there are a number of theoretical calculations based on different
schemes applied to constant and time-varying field configurations~\cite{GiesBook,PiazzaReview,Dinu2013}, 
to date there has been no direct experimental 
verification in pristine initial and final state conditions. 
The rapid development of high-intensity laser facilities,
such as the Extreme Light Infrastructure (ELI)~\cite{eli},
leads us to consider testing the propagation properties in focused
pump laser fields.
%starting from a setup with all optical laser lights. 
Once the calculation schemes have been tested in the context of QED, 
non-perturbative predictions can be reliably applied 
to more complicated intense fields: for instance, those in
strongly magnetized compact stars, such as magnetars~\cite{magnetar}, and
the early-stage of quark-gluon plasma accompanying thermal photons
in relativistic heavy-ion collisions~\cite{QGPgamma}, 
where interference between intense QED and 
intense quantum chromodynamic fields is expected~\cite{CME,QEDQCD}. 

The optical phase retardation $G$ between mutually orthogonal 
components of linearly polarized probe photon is given by
%\begin{equation}
$
G = 2\pi \Delta n  L \lambda^{-1},
$
%\end{equation}
where  $\lambda$ is the wave length of the probe photon,  $\Delta n$ is the relative refractive index change between the two  orthogonal components induced by the pump field and  
$L$ is the length of the birefringent region.

Several experiments have attempted to measure the magnetic birefringence of the vacuum \cite{pvlas, bmw, bfrt}.
For example, the PVLAS experiment~\cite{pvlas} can realize 
$\Delta n = 4 \times 10^{-23}[\mbox{T}^{-2}]\times (2.5[\mbox{T}])^2$
resulting in $G = 1.6 \times 10^{-9}$ with the total path length. 
This experiment utilizes the advantage of a static magnetic field 
to increase the phase shift using the long path length of the interaction region, 
which compensates for the smallness of $\Delta n$. 
 
In contrast to this approach utilizing a long $L$, 
we may consider combining a high-intensity pump laser and 
a high-energy probe to simultaneously increase $\Delta n$ and phase
retardation with a much shorter $\lambda$. 
Multi-petawatt class lasers have the capability of enhancing the relative refractive index change to $\Delta n \sim 10^{-11}$ 
at $\sim 10^{22}$~W/cm${}^2$~\cite{birefx3}. 
The use of X-ray probes was proposed~\cite{XLaser1,XLaser2} and its polarimetry technique exists~\cite{Xpol}.

In this paper we consider extending the probe energy up to the GeV regime.
The use of $\gamma$-ray probes to see the magnetic birefringent effect
has been proposed~\cite{GammaB1,GammaB2}.
What we propose here is to combine linearly polarized $\gamma$-ray probes
with the focused high-intensity laser field in order to realize $G \sim 1$.
Widening the probe energy range will allow complete measurement 
of the dispersion relation and enable accurate comparisons with 
the QED predictions.
On the other hand, we are required to newly develop a method to extract
phase retardation close to unity for the GeV probe.
The aim of this paper is to provide the concrete method to
determine it based on pair-wise topology of the Bethe-Heitler process,
{\it i.e.}, via the $\gamma$ to $e^+e^-$ conversion process in a polarimeter.

This paper consists of following sections.
In section II, we propose an experimental setup to probe
the laser-induced vacuum birefringence effect.
In section III, we discuss about the generation of highly linearly polarized 
probe $\gamma$-rays via nonlinear Compton scattering.
In section IV, we evaluate the amount of phase retardation 
by the QED effect with a parametrization of high-intensity laser pulse. 
In section V, we derive theoretical formulae
to parametrize phase retardation of probe photons
based on pair-wise topology of the Bethe-Heitler process in a polarimeter.
In section VI, we further provide a possible polarimeter design
and then evaluate the measurability of phase retardation
with a realistic set of laser parameters and statistics of pump laser pulses
by performing the detector simulation.
We finally conclude the realizability of the measurement and
discuss the prospect in section VII.
 
\section{A conceptual design of the proposed experiment}

\begin{figure}
\includegraphics[scale=0.42]{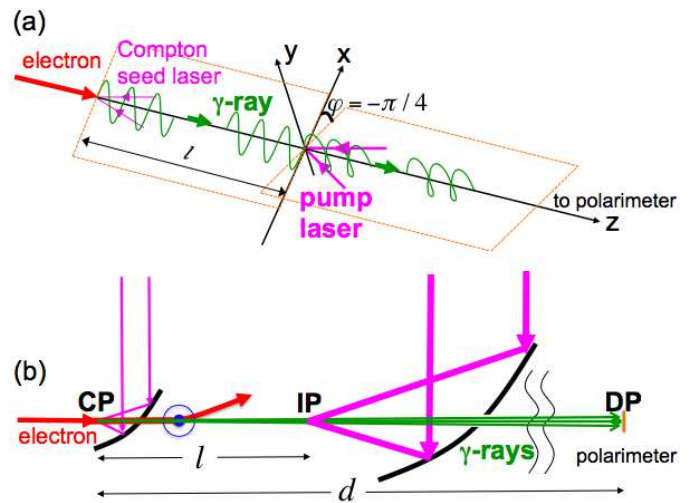}% Here is how to import EPS art
\caption{\label{Fig1} Conceptual experimental setup to 
investigate the laser-induced vacuum birefringence effect.
(a) Definitions of coordinates with respect to the linear polarization plane
of the Compton seed laser field and interaction points are provided.
(b) Colliding beam geometry with the alignment
of the incident electron beam, the parabola mirrors with pinholes
and a polarimeter. CP, IP, and DP indicate Compton scattering Point, 
Interaction Point, and Detection Point, respectively.
A distance $l$ between CP and IP and a distance $d$ between CP and DP
are introduced. The used electrons are bent in advance of
crossing with a pumping laser pulse at IP.}
\end{figure}

We consider a design of the experiment illustrated in Fig.\ref{Fig1} a)
for the measurement of the laser-induced birefringence effect.
Figure \ref{Fig1} b) shows colliding beam geometry with the alignment
of the incident electron beam, the parabola mirrors with pinholes
and the polarimeter.
The first mirror focuses a weaker laser pulse at CP to generate highly linearly
polarized probe $\gamma$-rays via Compton scattering with incident
monochromatic electron bunches and then
the second mirror focuses an intense laser pulse at IP to pump the vacuum
which is synchronized with the weaker laser pulse.
The probe $\gamma$-rays penetrate through the pumped domain and the polarization
states are altered by the vacuum birefringence effect.
Phase retardation embedded within the pumped domain is 
extracted from the pair-wise topology of the Bethe-Heitler process
within the polarimeter at DP. We introduce a distance $l$ between
CP and IP and a distance $d$ between CP and DP for later convenience.
We must require that the probe $\gamma$-ray energy is not too high 
in order to avoid the tunneling electron-positron pair production 
in the intense pump field. 
We assume 1 GeV probe $\gamma$-rays in the following design.
We simultaneously need a high degree of linear polarization for the probe 
$\gamma$-ray. 
We note that the reference polarization plane must be parallel to
the direction of the polarization 
of the Compton seed laser ($x-y$ plane in Fig.\ref{Fig1} (a)) 
and the pump laser must be aligned with
a relative rotation angle of $\pm \pi/4$ from that reference plane.
This configuration produces equal amplitudes for mutually orthogonal 
electromagnetic field components of probe photons and, hence, 
maximizes the visibility of phase retardation.

Given a 10 PW-class laser, for instance, what is available at the ELI project
with a typical wavelength of 800 nm, we can expect the completely synchronized
weak Compton seed and intense pump laser pulses as well as accelerated
unpolarized electrons by exploiting the laser-plasma acceleration 
technique~\cite{lpa1, lpa2}. We will assume that 5~GeV electrons
collide with the Compton seed laser pulses head-on.
The assumed electron energy is reasonable given the successful 
demonstration of quasi-monoenergetic electrons at 4.2 GeV \cite{acc1} 
with laser-plasma acceleration.
Of course, we may also use accurate 5~GeV electrons from a conventional 
accelerator as long as the electron source is synchronized with the
10 PW-class laser.
Based on this design, we discuss the individual elements from the upper stream
in Fig.\ref{Fig1} in the following sections.

\section{Generation of linearly polarized probe $\gamma$-rays}
As illustrated in Fig.\ref{Fig1},
linearly polarized probe $\gamma$-rays can be obtained by the inverse
Compton scattering in the forward region of incident electrons
interacting with linearly polarized laser pulses in head-on geometry.
In order to efficiently get higher energy photons for a given electron energy
and keep the spot size of generated $\gamma$-rays as small as 
possible~\cite{GammaGamma}, we utilize the multi-photon absorption 
in the nonlinear Compton scattering process. 
The $\gamma$-ray yields are estimated by using the cross sections
of the nonlinear Compton scattering process for the linearly polarized cases
parallel($\parallel$) and perpendicular($\perp$) to 
the linear polarization plane of the seed Compton laser field~\cite{slac1}.
As follows,
the differential cross sections for $n$-photon absorption are expressed
as a function of $u \equiv (k_1 k_2)/(k_1 p_2)$ 
with, respectively, the initial and final state photon four-momenta $k_1$ 
and $k_2$ and the final state electron four-momentum $p_2$, 
and of azimuthal angle $\phi$, which is defined as a 
rotation angle of the linear polarization plane of $k_2$ 
with respect to the incident linear polarization plane of $k_1$:
\begin{multline}\label{eq1cp}
\frac{d\sigma_{\parallel}}{du d\phi}=2{r_0}^2\frac{m^2}{s-m^2}\frac{1}{\eta^2(1+u)^2} 
\times \\
\left[ 
-2A^2_0\sigma + 4\eta^2\left(1+\frac{u^2}{4(1+u)}\right)(A^2_1-A_0A_2)
\right],
\end{multline}
\begin{multline}\label{eq2cp}
\frac{d\sigma_{\perp}}{du d\phi}=2{r_0}^2\frac{m^2}{s-m^2}\frac{1}{\eta^2(1+u)^2} 
\times \\
\left[ 
-2A^2_0(1-\sigma) + \eta^2\frac{u^2}{1+u}(A^2_1-A_0A_2)
\right]
\end{multline}
with
\beq\label{eq3cp}
\sigma \equiv 1 + \frac{(u_n-u)(\eta^2+1)(1-\cos2\phi)}{2u}
\eeq
where $A_l$($l$=0,1,2) are defined as
\begin{multline}\label{eq4cp}
A_l(\alpha_{Lin},\beta_{Lin},n) = \frac{1}{2\pi} \int^{\pi}_{-\pi} d\Phi \times \\
\cos^{(l)}\Phi \exp\{i\alpha_{Lin}\sin\Phi - i\beta_{Lin}\sin 2\Phi-in\Phi\}
\end{multline}
with
\beq\label{eq5cp}
\alpha_{Lin} = -2\sqrt{2}n\frac{\sqrt{u(u_n-u)}}{u_n}\frac{\eta}{\sqrt{1+\eta^2}}\cos\phi
\eeq
and
\beq\label{eq6cp}
\beta_{Lin} = \frac{n u}{2u_n}\frac{\eta^2}{1+\eta^2},
\eeq
where $s \equiv (p_1 + k_1)^2$ with 
the incident electron four-momentum $p_1$, the electron mass $m$, and 
$r_0 = e^2/(4\pi m) = \alpha/m = 2.82 \times 10^{-13} \mbox{cm}$
with the fine structure constant $\alpha = 1/137$.
These cross sections are characterized by a nonlinearity parameter
$\eta \equiv e\sqrt{-\langle A_{\mu} A^{\mu}\rangle}/mc^2$
with the four-vector potential of the incident photon $A_{\mu}$
accompanying a variable $u_n \equiv 2(k_1 p_1)n/(m^2(1+\eta^2))$
for the $n$-photon absorption case. 

We summarize a set of reachable beam parameters for 
a laser pulse, an electron bunch, and generated $\gamma$-ray 
probes per laser-electron crossing via nonlinear Compton scattering 
in Tab.\ref{Tab1}.
The laser power and intensity results in $\eta = 0.62$. 
A similar range of $\eta =0.4$ has been tested by the SLAC 
experiment~\cite{slac2} and we expect the cross sections are still valid.
Probe photons at GeV energies are generated within a small
scattering angle $\vartheta$ measured from the incident electron direction
and the dominant photon yield are confined in $\vartheta < 1/\gamma_{e}$ rad 
where $\gamma_{e}$ is the Lorentz factor of the incident electrons.
We consider a narrow bandwidth $\gamma$-rays within 1.015 - 1.021~GeV and 
emission angle less than or equal to 1/(10$\gamma_{e})  = 1.022 \times 10^{-5}$ 
rad. This energy range corresponds to the case when the number of absorbed 
laser photons reaches $n=3$. The limitation of the emission angle is actually 
necessary to select highly polarized $\gamma$-rays and can be required
by putting a narrow collimater at a distant point in front of the polarimeter.
\begin{table}
\begin{center}
\begin{tabular}{lc}
\hline
Laser wavelength & $\lambda = 800$ nm \\
Laser pulse energy & $E_L=2.2$ mJ \\
The number of laser photons & $N_L \equiv \frac{E_L}{hc / \lambda}$ \\
Laser pulse duration & $\tau_L$ = 33.5 fs \\
Laser pulse waist & ${w_L}_x = {w_L}_y = 1.6$ $\mu$m \\
Laser pulse power & 66.6 GW \\
Laser pulse intensity & 8.28 W/cm${}^2$ \\
\hline
Electron energy & 5 GeV \\
Electron bunch waist & ${w_e}_x = {w_e}_y = 1$ $\mu$m \\
Electron bunch length & 3 $\mu$m \\
\# of electrons & $N_e = 10^{10}$ \\
\hline
\# of absorbed laser photons & $n=3$ \\
$\gamma$-ray energy range in $\vartheta < 1/(10\gamma_e)$ & $1.015-1.021$~GeV\\
\# of $\gamma$-rays ($\parallel$) in $\vartheta < 1/(10\gamma_e)$ & ${N_{\gamma}}_{\parallel} = 64090$ \\
\# of $\gamma$-rays ($\perp$) in $\vartheta < 1/(10\gamma_e)$ & ${N_{\gamma}}_{\perp} = 890$ \\
\hline
\end{tabular}
\end{center}
\caption{A set of beam parameters for a laser pulse, an electron
bunch, and generated $\gamma$-ray probes per laser-electron crossing
via nonlinear Compton scattering.}
\label{Tab1}
\end{table}
The number of generated probe photons $N_{\gamma}$ per laser-electron
crossing can be numerically evaluated as 
\beq\label{eq7cp}
N_{\gamma} = \int^{\tau_L}_0 dt {\cal L} 
\int^{2\pi}_{0} d\phi \int^{1/(10\gamma_e)}_{0} 
d\vartheta \frac{d\sigma_{\parallel / \perp}}{du} \frac{du}{d\vartheta},
\eeq
where
${\cal L}$ is laser-electron luminosity per crossing in head-on collision 
geometry which is defined as
\beq\label{eq8cp}
{\cal L} = \frac{1}{\tau_L} \frac{N_e N_L}
{2\pi\sqrt{{w_L}^2_x + {w_e}^2_x} \sqrt{{w_L}^2_y + {w_e}^2_y}
}.
\eeq
The partially integrated cross sections are
$1.59 \times 10^{-20} \mu$m${}^2$ and $2.21 \times 10^{-22} \mu$m${}^2$
yielding the numbers of generated $\gamma$-rays 64090 and 890
for $\parallel$ and $\perp$ cases, respectively. 
% This results in the degree of linear polarization $P_l = 0.97$.

\begin{figure}
%\begin{center}
%\begin{tabular}{c}
%	\begin{minipage}{0.5\hsize}
	%\begin{center}
%	\includegraphics[scale=0.24]{Gamma_ray_Production.eps}
	%\end{center}
%	\end{minipage}
%	\begin{minipage}{0.5\hsize}
	%\begin{center}
%	\includegraphics[scale=0.24]{Gamma_ray_Polarization.eps}	
	%\end{center}
%	\end{minipage}
%\end{tabular}
\includegraphics[scale=0.48]{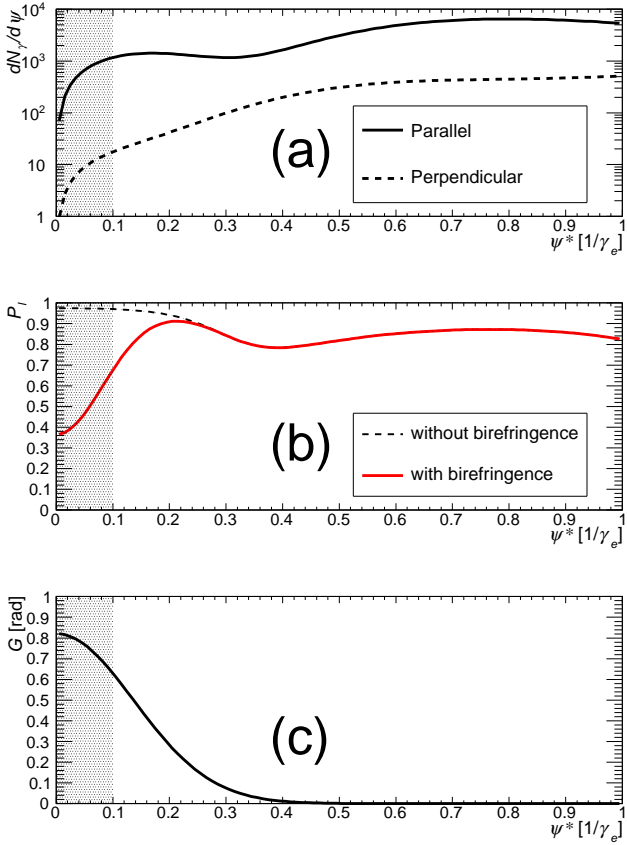}
\caption{\label{Fig2} (a) Incident $\gamma$-ray yields as a function of the emission angle for orthogonal linear polarization states, respectively.
(b) The degree of linear polarization of $\gamma$-rays. The polarization-flip effect due to the laser-induced birefringence depends on the emission angle.
(c) The corresponding phase retardation $G$.}
%\end{center}
\end{figure}

\section{Expected phase retardation in the conceptual design}
In the quantum mechanical view point,
phase retardation may be interpreted 
as a consequence of polarization flips of probe photons.
The degree of linear polarization of probe photons can be defined as
\begin{equation}\label{eq_pl}
P_{l} = \frac{N_{\parallel} - N_{\perp}}{N_{\parallel} + N_{\perp}},
\end{equation}
where $N_{\parallel}$ and $N_{\perp}$ are, respectively, the numbers of probe
photons with linear polarization states parallel and perpendicular
to the direction of the linear polarization of the pump laser.
The polarization-flip phenomenon in an intense pump field has been discussed 
and quantified by Dinu et al.~\cite{Dinu2014}.
In order to parametrize the flipping probability of probe photons 
with energy $\omega$ for a given pump laser pulse 
as summarized in Tab.\ref{Tab1}, we impose following requirements:
\begin{itemize}
\item bandwidth is small: $(\Delta\omega_0/\omega_0)^2 \ll 1$
\item validate pulse approximation: $(s^2\omega_0/(4\Delta\omega_0))^2 \ll 1$
\item Heisenberg-Euler limit: $2\omega\omega_0/m^2 \ll 1$,
\end{itemize}
where $\omega_0$ is the pump photon energy, $\Delta\omega_0$ satisfies
$\tau_0 = \sqrt{8\log2}/\Delta\omega_0$ with the pump pulse duration $\tau_0$, 
and $s = \lambda_0/(\pi w_0)$ corresponds to the beam divergence
with the pump beam waist $w_0$ and the pump laser wavelength $\lambda_0$.
Under these conditions, 
suppose probe photons collide head-on with a focused pump pulse
having a Gaussian profile, as illustrated in Fig.\ref{Fig1}, 
the flipping probability of probe photons is approximated as
\begin{eqnarray}
P_f = \left( \frac{\alpha}{15} \frac{1}{E_{s}^{2}} \frac{ {\cal E}_0 \omega}{\pi^{2}w_{0}^{2}}\right) ^{2} {\rm e}^{-\frac{4\psi^{2}}{\psi_{0}^{2}}} \nonumber \\
\mbox{with} \quad \psi_{0} = {\rm tan}^{-1} \frac{w_{0}}{l}, \label{eq:pflip}
\end{eqnarray}
where $\alpha$ is the fine structure constant, 
$E_{s}$ is the Schwinger critical electric field, 
$w_{0}$ is the waist size of the focused laser, 
${\cal E}_0$ is the energy of the pump laser, 
$\psi$ is the incident angle of probe $\gamma$-rays 
with respect to the head-on direction and 
$l$ is the distance between CP and IP in Fig.\ref{Fig1} (b). 
The effect of the incident angle distribution of $\gamma$-rays or 
misalignment with respect to the head-on collisions is 
expressed as the exponential reduction of the flipping probability. 

A natural experimental observable is thus the reduction of 
the degree of linear polarization from $P_{l}$ to $(1-2P_f)P_l$.
However, in the case of GeV probe photons, there is no known polarizer
to directly determine $N_{\parallel}$ and $N_{\perp}$ in experiments.
In addition, there could be other sources to reduce $P_l$ than the pure phase 
retardation effect embedded at IP in Fig.\ref{Fig1} (b). 
Dinu et al.~\cite{Dinu2014} also discuss the relation between the flipping
probability and the phase retardation.
If we have a way to directly determine the phase retardation itself
in a polarimeter, we may be able to discriminate the true birefringent effect
from the other sources of reduction of $P_l$. In the next section,
we will provide the theoretical basis for this idea.
We thus provide here the definition of phase retardation $G$ 
in our notation in accordance with the following section,
\beq
G \equiv 2\delta \sim \frac{2\sqrt{P_f}}{1+\bar{w}}
e^{2\rho^2\frac{\bar{w}}{1+\bar{w}}}
\eeq
where $\delta$ is the same definition as in Eq.(36) of 
Dinu et al.'s paper~\cite{Dinu2014} as a function of $\rho \equiv r/w_0$ 
with the transverse position $r$ relative to the beam waist $w_0$
by introducing the ratio of probe to target
beam waists $\bar{w} \equiv 2w^2/w^2_0$.

We now consider the case summarized in Tab.\ref{Tab2} where
the generated polarized $\gamma$-rays penetrate through
the focal region of the pump laser after traveling a distance $l$.
Assuming a conservative waist size for the focal spot of $w_0 = 2.4$ ${\rm \mu m}$,  a wavelength of 800 nm, a pulse energy of 200 J and an intensity of 
$3.7\times10^{22}$ W/cm$^{2}$,
the degree of linear polarization of the incident $\gamma$-rays is 
expected to change from $\langle P_{l} \rangle = 0.97$ to
 $\langle (1-2P_f) P_l \rangle = 0.53$ after passing through the laser-induced birefringent vacuum, where $\langle \cdot \cdot \cdot \rangle$ represents the weighted mean of the degree of the polarization over the angular range from $\psi$ = 0 to $\psi$ = $1/(10\gamma_{e})$ and 
the flipping probability, $P_f$, has been calculated 
with the given laser parameters and $l$ = 10 cm.
\begin{table}
\begin{center}
\begin{tabular}{lc}
\hline
Pump laser photon energy & $\omega_0=1.55$~eV\\
Pump laser pulse energy & ${\cal E}_0 = 200$ J \\
Pump laser pulse duration & $\tau_0 = 30$ fs \\
Pump laser pulse beam diameter & 50 cm \\
Pump laser pulse waist with F\#=2.35 & $w_0 = 2.4$ $\mu$m\\
Pump laser pulse intensity & $3.7 \times 10^{22}$~W/cm${}^2$ \\
Probe $\gamma$-ray energy & $\omega = 1$ GeV \\
Distance $l$ & 10 cm \\
Probe $\gamma$-ray waist, $l\tan\{1/(10\gamma_e)\}$, at IP & $w=1.0$~$\mu$m\\
Geometrically averaged phase retardation & $\langle G \rangle = 0.72$ \\
\hline
\end{tabular}
\end{center}
\caption{A set of parameters for a single pump laser pulse and
linearly polarized 1 GeV probe photons.}
\label{Tab2}
\end{table}
Figure \ref{Fig2} (a) shows incident $\gamma$-ray yields
as a function of the emission angle with respect to 
the direction of the incident electron based on 
Eq.(\ref{eq1cp}), (\ref{eq2cp}), (\ref{eq7cp}) and (\ref{eq8cp}).
The horizontal axis $\psi^{*}$ in Fig. \ref{Fig2} is the deduced 
emission angle, which is normalized to the inverse of 
the Lorentz factor for 5 GeV electrons.
The components parallel and perpendicular to the polarization direction 
of the Compton seed laser are depicted by solid and dotted lines
in Fig.\ref{Fig2} (a), respectively. 
The polarization-flip effect appears as a reduction of the degree of 
linear polarization, as shown in Fig.\ref{Fig2} (b). 
A large polarization-flip effect is visible in the forward direction, 
especially for $\psi^* < 1/10$. 
Figure \ref{Fig2} (c) shows the corresponding phase retardation $G$.
The average of $G$ within $\psi^{*} < 1/10$ with $\bar{w} = 0.35$ reaches 
$\langle G \rangle = 0.72$ rad.

\section{Extracting phase retardation from pair-wise topology of the
Bethe-Heitler process}
Determining of the degree of linear polarization of incident photons 
via the Bethe-Heitler process is proposed in Ref.\cite{pair_pol1}
and the method has been applied to several experiments, for example,
Ref.\cite{pair_pol5,pair_pol6,pair_pol7}.
The detailed theoretical basis can also be found in 
Ref.\cite{pair_pol3,pair_pol4}.
However, there is no explicit calculation for a general ellipsoidally
polarized case to date. 
Because phase retardation $G$ is close to unity in our case,
we cannot approximate the polarization state of $\gamma$-rays 
penetrating though the pumped domain as the linearly polarized state anymore.
In this section we thus derive the pair-wise angular distribution 
with contemporary notations, {\it e.g.}, found in Ref.\cite{Greiner}
in order to explicitly implement phase retardation $G$ into polarization vectors
of incident photons so that the theoretical functional form is directly 
applicable to the concrete polarimeter proposed in the next section.

The differential cross section $d\sigma$ is expressed as
\beq\label{eq1}
d\sigma = \int \frac{|S_{fi}|^2}{T\frac{v_{in}}{V}} V
\frac{d^3p_+}{(2\pi)^3} V \frac{d^3p_-}{(2\pi)^3},
%=
%\int \frac{|S_{fi}|^2}{T\frac{1}{V}} V
%\frac{d^3p_+}{(2\pi)^3} V \frac{d^3p_-}{(2\pi)^3}
\eeq
where $S_{fi}$ is the transition amplitude within a time interval $T$
and a normalized volume $V$ of the conversion process
from an initial photon state $i$ with relative velocity $v_{in}=c$
to a fixed Coulomb potential of a target nucleus 
into an electron and positron pair in the final state $f$ whose
four-momenta are $p_-$ and $p_+$, respectively.
With respect to the static Coulomb potential with a point charge $-Ze$,
the transition amplitude is described as~\cite{Greiner}
\begin{multline}\label{eq2}
S_{fi} = Ze^32\pi\delta(E_+ + E_- - \omega)\sqrt{\frac{4\pi}{2\omega V}}\sqrt{\frac{m^2}{E_+ E_- V^2}}\frac{4\pi}{|q|^2} \\
\times \bar{u}(p_-,s_-) \left[
(-i\s{\epsilon})\frac{i}{\s{p_-}+\s{k}-m}(-i\gamma^0) + 
\right. \\ \left.
(-i\gamma^0)\frac{i}{-\s{p_+}+\s{k}-m}(-i\s{\epsilon})
\right] v(p_+,s_+) \\
\equiv -i Ze^32\pi\delta(E_+ + E_- - \omega)\sqrt{\frac{4\pi}{2\omega V}}\sqrt{\frac{m^2}{E_+ E_- V^2}}\frac{4\pi}{|q|^2} \\
\times \bar{u}(p_-,s_-)\Gamma v(p_+,s_+)
% \epsilon^{\mu}M_{\mu}
\end{multline}
where 
electron and positron spinors, $u$ and $v$, respectively,
with the equal mass $m$,
Dirac matrices $\gamma^{\mu}$ with $\mu=0\sim3$ giving the Feynman slash
notation $\s{A} \equiv \gamma^{\mu} A_{\mu}$ for an arbitrary 
four-dimensional vector $A$,
incident photon four-momentum $k\equiv(\omega,\vec{k})$ with the
four-dimensional polarization vector $\epsilon$,
four-momentum transfer $q \equiv p_+ + p_- - k$ with
$p_+\equiv(E_+,\vec{p_+})$ and $p_-\equiv(E_-,\vec{p_-})$,
and $\Gamma$ is defined in Eq.(\ref{eq5}).
With $\alpha = e^2/(\hbar c) \equiv e^2$ and, in general,
$(2\pi\delta^2(E_f-E_i))^2=2\pi\delta(0)2\pi\delta(E_f-E_i)=2\pi T\delta(E_f-E_i)$, we can express the square of the transition amplitude as
\begin{multline}\label{eq3}
|S_{fi}|^2 = Z^2e^6(2\pi)^2\delta^2(E_+ + E_- - \omega) \times \\
\left(\frac{4\pi}{2\omega V}\right)
\left(\frac{m^2}{E_+ E_- V^2}\right)
\frac{(4\pi)^2}{|q|^4} \cal{F} \\
= Z^2\alpha^3 2\pi T \delta(E_+ + E_- - \omega) \times \qquad \qquad \\
(4\pi)^3 \frac{1}{V^3} \frac{m^2}{2\omega E_+ E_-} \frac{1}{|q|^4} \cal{F}
\end{multline}
with
\begin{multline}\label{eq4}
{\cal F} \equiv
\left( \bar{u}(p_-,s_-)\Gamma v(p_+,s_+) \right)
\left( \bar{u}(p_-,s_-)\Gamma v(p_+,s_+) \right)^{\dagger} \\
=
\left( \bar{u}(p_-,s_-)\Gamma v(p_+,s_+) \right)
\left( \bar{v}(p_+,s_+)\bar{\Gamma} u(p_-,s_-) \right)
\end{multline}
where
\beq\label{eq5}
\Gamma \equiv
\s{\epsilon}\frac{\s{p_-}-\s{k}+m}{-2p_-\cdot k}\gamma^0
+\gamma^0\frac{-\s{p_+}+\s{k}+m}{-2p_+\cdot k}\s{\epsilon}
\eeq
and
\beq\label{eq6}
\bar{\Gamma} \equiv \gamma^0 \Gamma^{\dagger} \gamma^0 =
\gamma^0\frac{\s{p_-}-\s{k}+m}{-2p_-\cdot k}\s{\epsilon}^*
+\s{\epsilon}^*\frac{-\s{p_+}+\s{k}+m}{-2p_+\cdot k}\gamma^0.
\eeq
Here we note that $*$ is explicitly displayed only for the polarization vector
part including imaginary components as we discuss later.
Substituting Eq.(\ref{eq3}) into Eq.(\ref{eq1})
with $v_{in}=c \equiv 1$, we get
\begin{multline}\label{eq7}
d\sigma = \int 8 Z^2\alpha^3 \delta(E_+ + E_- - \omega)(2\pi)^4 
\times\\
\frac{m^2}{2\omega E_+ E_-} \frac{1}{|q|^4} {\cal F}
\frac{d^3p_+ d^3p_-}{(2\pi)^6} \\
=
\int \frac{4 Z^2\alpha^3 m^2}{(2\pi)^2 \omega E_+ E_-} \frac{1}{|q|^4}
\delta(E_+ + E_- - \omega) d^3p_+ d^3p_- {\cal F} \qquad \\
=
\int dE_- \frac{4 Z^2\alpha^3 m^2}{(2\pi)^2 \omega E_+ E_-} \frac{1}{|q|^4}
\delta(E_+ + E_- - \omega)
\times \mbox{\hspace{3.7cm}} \\
|\vec{p_-}|E_- d\Omega_- |\vec{p_+}|E_+ dE_+ d\Omega_+ {\cal F} \\
=
\frac{Z^2\alpha^3}{(2\pi)^2} \frac{4m^2}{\omega |q|^4}
|\vec{p_-}||\vec{p_+}| dE_+ d\Omega_+ d\Omega_-
\Theta(\omega - E_+ - m) {\cal F} \quad
\end{multline}

We then define $F$ by summing ${\cal F}$ over the possible
electron and positron spin states as
\begin{multline}\label{eq8}
F \equiv
\sum_{s_-,s_+}
\text{tr}\left[\frac{\s{p_-}+m}{2m}\Gamma\frac{\s{p_+}-m}{2m}\bar{\Gamma}\right]
\\ \equiv
\frac{1}{16m^2} \left[
\frac{A}{(p_-\cdot k)^2}
+
\frac{B}{(p_-\cdot k)(p_+\cdot k)}
+ \right. \\ \left.
\frac{C}{(p_+\cdot k)(p_-\cdot k)}
+
\frac{D}{(p_+\cdot k)^2}
\right]
\end{multline}
with
\begin{multline}
A \equiv \text{tr}\left[
(\s{p_-}+m)\s{\epsilon}(\s{p_-}-\s{k}+m)\gamma^0 \right. \\ \left.
(\s{p_+}-m)\gamma^0(\s{p_-}-\s{k}+m)\s{\epsilon}^*
\right] \\
B \equiv \text{tr}\left[
(\s{p_-}+m)\s{\epsilon}(\s{p_-}-\s{k}+m)\gamma^0 \qquad\qquad\qquad\qquad \right. \\ \left.
(\s{p_+}-m)\s{\epsilon}^*(\s{p_-}-\s{k}+m)\gamma^0
\right]
\\
C \equiv \text{tr}\left[
(\s{p_-}+m)\gamma^0(-\s{p_+}+\s{k}+m)\s{\epsilon} \qquad\qquad\qquad\quad \right. \\ \left.
(\s{p_+}-m)\gamma^0(\s{p_-}-\s{k}+m)\s{\epsilon}^*
\right]
\\
D \equiv \text{tr}\left[
(\s{p_-}+m)\gamma^0(-\s{p_+}+\s{k}+m)\s{\epsilon} \qquad\qquad\qquad\quad \right. \\ \left.
(\s{p_+}-m)\s{\epsilon}^*(-\s{p_+}+\s{k}+m)\gamma^0
\right] \notag
\end{multline}
where we note $F$ has dimension of eV${}^{-2}$.

Let us remind of the Jones matrix in order to introduce a general
ellipsoidally polarized vector beginning from a linearly polarized
photon in the $x$-direction. 
The Jones matrix in $x-y$ coordinate is defined as~\cite{Yariv}
\beqa
W(\varphi, G) = 
\left(
    \begin{array}{cc}
      \cos\varphi & \sin\varphi \\
      -\sin\varphi & \cos\varphi \\
    \end{array}
\right)
\times \nnb \\
\left(
    \begin{array}{cc}
      e^{-iG/2} & 0 \\
      0 & e^{iG/2} \\
    \end{array}
\right)
\left(
    \begin{array}{cc}
      \cos\varphi & -\sin\varphi \\
      \sin\varphi & \cos\varphi \\
    \end{array}
\right),
\eeqa
where $\varphi$ denotes a rotation angle of the linear
polarization plane of the pump laser field with respect
to the linear polarization plane of an incident probe $\gamma$-ray
in our case.  
Following Fig.\ref{Fig1} (a) indicating $\varphi = -\pi/4$ with respect to 
the $x$-axis, the $x-y$ polarization vector after penetrating through
the pumped domain is expressed as 
\beqa
\left(
    \begin{array}{c}
      \epsilon_x \\
      \epsilon_y \\
    \end{array}
\right)
= W(-\pi/4, G)
\left(
    \begin{array}{c}
      1 \\
      0 \\
    \end{array}
\right)
=
\left(
    \begin{array}{c}
      \cos\frac{G}{2} \\
     i\sin\frac{G}{2} \\
    \end{array}
\right).
\eeqa
We then extend this ellipsoidally polarized vector of a probe $\gamma$-ray 
into a four-dimensional polarization vector as follows
\beq\label{eq9}
\epsilon \equiv 
\cos\frac{G}{2}g_1 + i\sin\frac{G}{2}g_2
\eeq
with $g_1 = (0,1,0,0)$ and $g_2=(0,0,1,0)$. 
In this case the corresponding Feynman slash variables are defined as
\begin{gather}\label{eq10}
\s{\epsilon} = \cos\frac{G}{2}\gamma^{1} + i\sin\frac{G}{2}\gamma^{2}
\\ \notag
\s{\epsilon}^* = \cos\frac{G}{2}\gamma^{1} - i\sin\frac{G}{2}\gamma^{2}.
\end{gather}
By performing the trace calculation in Eq.(\ref{eq8}) with $g_0= (1,0,0,0)$,
$F$ can be expressed with products of four-vectors as follows:
\begin{multline}\label{eqF}
F = \frac{1}{16m^2}\frac{8}{(k \cdot p_-)^2(k \cdot p_+)^2}
\Biggl[
 (k \cdot p_+)
 \biggl\{
  (k \cdot p_-)
   \\ \times
   \Bigl(
    (k \cdot p_+)
    \bigl\{
     (\cos G + 1)(g_1 \cdot p_-)^2 - (\cos G -1)(g_2 \cdot p_-)^2      
     \\
     + 2\omega\bigl((g_0 \cdot p_-) + (g_0 \cdot p_+)\bigr) - (k \cdot p_+)
    \bigr\}
    -2\omega^2(p_- \cdot p_+) - 2m^2\omega^2
   \Bigr)
   \\
   -(k \cdot p_+)
   \Bigl(
    2(g_0 \cdot p_+) \bigl( \omega - (g_0 \cdot p_-) \bigr)
    -(k \cdot p_+) + (p_- \cdot p_+) + m^2
   \Bigr)
   \\ \times
   \Bigl(
    (\cos G + 1)(g_1 \cdot p_-)^2 - (\cos G -1)(g_2 \cdot p_-)^2
   \Bigr)
   \\
   + 2\omega
   \Bigl(
    (g_0 \cdot p_-) + (g_0 \cdot p_+)
   \Bigr) (k \cdot p_-)^2 - (k \cdot p_-)^3
 \biggr\}
%%%%%%%%%%%
 \\
 + (\cos G + 1)(g_1 \cdot p_+)^2(k \cdot p_-)^2
   \\ \times
   \bigl\{ 
    2(g_0 \cdot p_-) \bigl( (g_0 \cdot p_+) - \omega \bigr)
    + (k \cdot p_-) + (k \cdot p_+) - (p_- \cdot p_+) - m^2
   \bigr\}
   \\
 + (\cos G - 1)(g_2 \cdot p_+)^2(k \cdot p_-)^2
   \\ \times
   \bigl\{ 
    2(g_0 \cdot p_-) \bigl( \omega - (g_0 \cdot p_+) \bigr)
    - (k \cdot p_-) - (k \cdot p_+) + (p_- \cdot p_+) + m^2
   \bigr\}
   \\
 + 2(\cos G + 1)(g_1 \cdot p_-) (g_1 \cdot p_+) (k \cdot p_-) (k \cdot p_+)
   \\ \times
   \bigl\{ 
    (g_0 \cdot p_-) \bigl( \omega - 2(g_0 \cdot p_+) \bigr)
    + \omega(g_0 \cdot p_+) - (k \cdot p_-) \\
    - (k \cdot p_+) + (p_- \cdot p_+) + m^2 - \omega^2
   \bigr\}
   \\
 + 2(\cos G - 1)(g_2 \cdot p_-) (g_2 \cdot p_+) (k \cdot p_-) (k \cdot p_+)
   \\ \times
   \bigl\{
    (g_0 \cdot p_-) \bigl( 2(g_0 \cdot p_+) - \omega \bigr)
    - \omega(g_0 \cdot p_+) + (k \cdot p_-) \\
    + (k \cdot p_+) - (p_- \cdot p_+) - m^2 + \omega^2
   \bigr\}
\Biggr] - 16.
\end{multline}
We used FeynmanCalc\cite{FeynmanCalc} for the trace calculation.
In the proceeding calculations we introduce 
following definitions of four-momentum with components 
in Cartesian coordinates and also polar coordinates for $p_+$ and $p_-$:
\begin{multline}\label{eq11}
k \equiv (\omega,0,0,\omega) \\
p_+ \equiv (E_+,{p_+}_x,{p_+}_y,{p_+}_z) \mbox{\hspace{4.5cm}}\\
= (E_+,|\vec{p_+}|\sin\theta_+\cos\phi_+,|\vec{p_+}|\sin\theta_+\sin\phi_+,|\vec{p_+}|\cos\theta_+)\\
p_- \equiv (E_-,{p_-}_x,{p_-}_y,{p_-}_z) \mbox{\hspace{4.5cm}}\\
= (E_-,|\vec{p_-}|\sin\theta_-\cos\phi_-,|\vec{p_-}|\sin\theta_-\sin\phi_-,|\vec{p_-}|\cos\theta_-) \\
q \equiv (p_+ + p_-) - k \mbox{\hspace{5.0cm}}\\
= (0,\quad {p_+}_x+{p_-}_x,\quad {p_+}_y+{p_-}_y,\quad {p_+}_z+{p_-}_z-\omega),
\end{multline}
where energy conservation $\omega = E_+ + E_-$ is required for $q$.
Because the cross section is maximized in the case of $q \rightarrow 0$,
we consider only symmetrically emitted $e^+e^-$ pairs
within the same emission plane with following conditions
\beqa\label{eq12}
E_- &=& E_+ = \omega/2 \\\nnb
\theta_- &=& \theta_+ \\\nnb
\phi_- &=& \phi_+ + \pi.
\eeqa
Fortunately, in this symmetric case, the exact analytical expression
for the square of the invariant amplitude can be quite simplified as follows
\begin{multline}\label{eq13}
F_{sym} = \frac{1}{16m^2} \frac{16\left(\omega^2 - 4m^2\right)}{\omega^2} \sin^2\theta_+ (\cos G \cos2 \phi_+ + 1). \\
\end{multline}
From Eq.(\ref{eq7}) with
$|\vec{p_-}|=|\vec{p_+}| = \sqrt{\frac{\omega^2}{4}-m^2}$
and $dE_+ = \frac{1}{2}d\omega$, $d\theta_+ = d\theta_-$ and $d\phi_+ = d\phi_-$,
the differential cross section in terms of positron-relevant variables
for the symmetric case is expressed as
\begin{multline}\label{eq14}
\frac{d\sigma_{sym}}{d\omega d\phi_+ d\theta_+} = 
\int_0^{2\pi} d\phi_- \delta(\phi_- - (\phi_+ + \pi))
\int_0^{\pi} d\theta_- \delta(\theta_- - \theta_+) \\
\times \frac{1}{2}\frac{Z^2\alpha^3}{(2\pi)^2} \frac{4m^2}{\omega |\vec{q}|^4}
|\vec{p_+}|\sin\theta_+ |\vec{p_-}|\sin\theta_- F \Theta(\omega/2 - m) \\
= 
\frac{1}{2}\frac{Z^2\alpha^3}{(2\pi)^2} \frac{4m^2}{\omega |\vec{q}_{sym}|^4}
p^2_{+} \sin^2\theta_+ F_{sym} \Theta(\omega/2 - m) \mbox{\hspace{1.5cm}}\\
=
\frac{Z^2(2\alpha)^3}{(2\pi)^2\omega^3}
\left(\frac{p_{+}\sin\theta_+}{|\vec{q}_{sym}|}\right)^4 
(\cos G \cos2\phi_+ + 1) \Theta(\omega/2 - m),
\end{multline}
with
%block functions
%$\hat{\delta}(x) \equiv 
%\{\Theta(x+\epsilon/2)-\Theta(x-\epsilon/2)\}/\epsilon$ 
%equivalent to $\lim_{\epsilon \rightarrow 0} \hat{\delta}(x) = \delta(x)$
%are introduced in order to explicitly show that we indeed convolute probability
%distributions in the angular variables
%and
$|\vec{q}_{sym}|^4 \equiv \{2|\vec{p_+}|\cos\theta_+  - \omega\}^4$
via the relation 
$|\vec{q}|^2=-q^2 = \{ (|\vec{p_+}|\cos\theta_+ + |\vec{p_-}|\cos\theta_-) - \omega\}^2$.

We then express the partially integrated cross section within an experimental
coverage $0\le\theta_+\le\Delta\theta$ and $0\le\phi < 2\pi$ in a measurement. 
This quantity gives us the conversion efficiency into useful symmetric
pairs for the determination of phase retardation $G$.
We thus further integrate over $d\phi_+$, $d\theta_+$ and $d\omega$, which gives
\begin{multline}\label{eq18}
\sigma_{sym} \sim
\frac{4Z^2\alpha}{\pi} \left(\frac{\alpha}{m}\right)^2
\int^{\langle\omega\rangle+\Delta\omega}_{\langle\omega\rangle-\Delta\omega} d\omega 
\int_0^{\Delta\theta} d\theta_+ \times \\
\frac{m^2(\omega^2/4-m^2)^2\sin^4\theta_+}
{\omega^3(2\sqrt{\omega^2/4-m^2}\cos\theta_+ - \omega)^4} \Theta(\omega/2 - m) \\
\equiv
\frac{4Z^2\alpha}{\pi} \sigma_e H(\langle\omega\rangle,\Delta\omega,\Delta\theta)
\mbox{\hspace{2.7cm}}
\end{multline}
with $\sigma_e \equiv (\alpha/m)^2 = (2.8\times10^{-13}\text{cm})^2 = 0.0784~\text{b}$.

For $\langle\omega\rangle = 1$~GeV, $\Delta\omega/\langle\omega\rangle = 0.005$ and $\Delta\theta=0.01$,
$H(\langle\omega\rangle,\Delta\omega,\Delta\theta) = 0.239$ is obtained. If we choose
$Z=79$~(a gold converter), the cross section reaches $\sigma_{sym} = 1.1$~b, 
even if we require the special symmetric case of pair-wise topology in experiments.

\section{Polarimetry}
\subsection{Parametrization for pair-wise angular distributions}
The degree of linear polarization of the $\gamma$-rays
is characterized by the anisotropic angular distribution of 
emission planes containing electron-positron pairs 
with respect to the polarization plane of the incident $\gamma$-rays
~\cite{pair_pol1,pair_pol3,pair_pol4,pair_pol5,pair_pol6,pair_pol7}. 
At the same time energies of $\gamma$-rays above 100 MeV can be 
reconstructed by the kinematical relations for 
the conversion process from a $\gamma$-ray into an electron-positron pair. 
If offline selections in experiments allow us to impose the symmetric 
condition in Eq.(\ref{eq12}), we can parametrize the angular distribution
based on the expression in Eq.(\ref{eq14}) by taking other bias factors 
into account.
One of the biases would be initially caused by the degree of linear polarization
of probe $\gamma$-rays, because we have to accept a finite angular spread
of incident $\gamma$-rays as indicated in Fig.\ref{Fig2} (a).
Even if we limit the angular spread in $\psi^{*}<0.1$, the averaged
degree of linear polarization is 97 \%.
In such a case, by denoting $\phi$ as the angle of 
the $e^{+}e^{-}$ emission plane
with respect to the linear polarization plane of the Compton seed laser
for the case $G=0$,
linearly polarized photons in $\parallel$ and $\perp$ directions 
cause an uncorrelated statistical ensemble of 
$P\cos 2\phi$ and $Q\cos(2(\phi + \pi/2))$
with the different statistical weights 
$P=N_{\parallel}/(N_{\parallel}+N_{\perp})$ and 
$Q=N_{\perp}/(N_{\parallel}+N_{\perp})$ resulting in 
$P_l \cos(2\phi)$ with $P_l = P-Q$ as defined in Eq.(\ref{eq_pl}).
In addition to this known bias, in general,
experimental resolutions would reduce the amplitude of the modulation. 
By taking these factors into account,
the angular distribution of emission planes containing individual $e^{+}e^{-}$ 
pairs can be parametrized as follows 
\begin{equation}\label{eq_pol} 
\frac{dN_{e^+e^-}}{d\phi} =N_0 
\left( 1+A P_l \cos G \cos (2\phi-\phi_0) \right),
\end{equation}	
where $N_0$ is the number of $e^{+}e^{-}$ pairs in the unpolarized case
and $\phi_0$ is an offset phase. The analyzing power, $A$, refers to
the reduction of anisotropy caused by experimental resolution.
The offset phase is introduced to allow the offset angle of
the polarimeter plane ($x-z$ plane in Fig.\ref{Fig3})
with respect to the linear polarization plane of the Compton seed laser~
($x-z$ plane in Fig.\ref{Fig1} (a)). 
In the proceeding discussion, we always assume $\phi_0 = 0$.

\subsection{Polarimeter design}
There are two key issues in the design of the $\gamma$-ray polarimeter.
The first is how to deal with a large number of $\gamma$-rays 
confined within a cone angle of $1/(10\gamma_{e})$. In our estimation
summarized in Tab.\ref{Tab1}, $\sim 6.5 \times 10^{4}$ $\gamma$-rays at 1 GeV 
are expected to enter into a detector at a time.
For this purpose, the $\gamma$-ray converter must be carefully chosen 
in order to adjust the number of $e^{-}e^{+}$ pairs depending on 
the handling capability of the polarimeter.
To accurately spot the 1/(10$\gamma_{e}) \sim 10^{-5}$~rad, 
we need to locate the converter far from the interaction point. 
If the converter is located at $d=10$~m in Fig.\ref{Fig1}~(b), 
the angular spread results in the transverse 
spread of 100~$\mu$m on the converter.
%Because we will impose a fixed incident spot 
%in the track reconstruction algorithm without resolving
%individual incident points on the converter,
%the position resolution in
%a charged particle sensor beyond the incident point uncertainly of 10 $\mu$m 
%is not necessary. 
This suggests that a conventional silicon pixel type sensor with a few
10~$\mu$m resolution is useful for this type of polarimetry.

The second issue is how to accurately reconstruct an emission plane
based on the momentum vectors of an $e^{-}e^{+}$ pair produced at the conversion point,
which has the greatest effect on the analyzing power in the end.
The opening angle of a pair at the conversion point is the key information needed to correctly reconstruct the anisotropy in Eq.(\ref{eq_pol}). 
The original angle is, however, smeared by multiple Coulomb scatterings during passage through the conversion material.
%depending on the converter thickness. 
In addition, multiple Coulomb scatterings inside each pixel sensor also give rise to a displacement of the measured hits from the ideal trajectory of a charged particle.
This displacement degrades the track finding and reconstructing capabilities and reduces the analyzing power.      
%the track finding and reconstruction capabilities are also degraded due to Multiple Coulomb scatterings inside the 
%pixel sensors. 
Thus, the thickness of the detector materials must be controlled 
to keep the analyzing power at an acceptable level.

\begin{figure}
\begin{center}
%\begin{tabular}{c}
	%\begin{minipage}{0.5\hsize}
	%\begin{center}
	\includegraphics[scale=0.22]{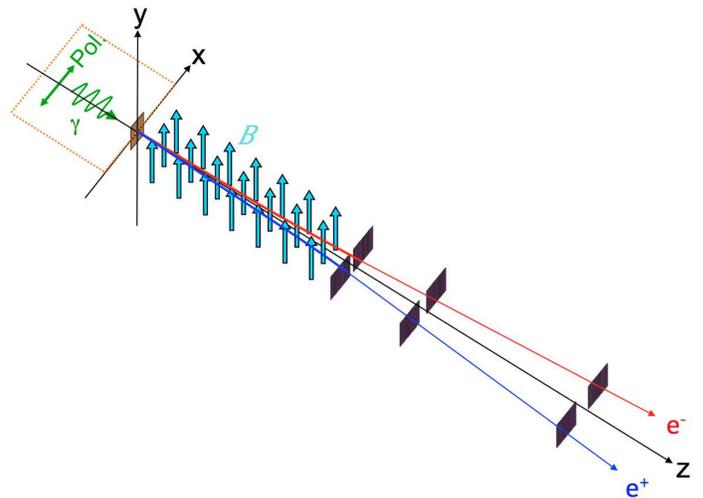}
	%\end{center}
	%\end{minipage}
	%\begin{minipage}{0.5\hsize}
	%\begin{center}
	%\includegraphics[scale=0.145]{GPC_Event_Display.eps}	
		%\end{center}
	%\end{minipage}
%\end{tabular}
\caption{\label{Fig3} Configuration of the detection elements
for $\gamma$-ray polarimetry and spectroscopy via the Bethe-Heitler process
on a thin converter located at the origin.
}
\end{center}
\end{figure}

The minimum elements of the detector design are illustrated in Fig.\ref{Fig3}.
The detection system simultaneously performs spectroscopy and polarimetry 
for a multiple $\gamma$-ray injection.
It is composed of a converter at the front followed by a narrow collimator
to guarantee the narrow angular spread, that is, narrow energy band of the
incident probe $\gamma$-rays, 
a static magnetic field, and three-layers of pixel sensors. 
The converter is chosen to suppress the smearing effect due to 
multiple Coulomb scatterings but to keep the pair creation efficiency 
reasonably high. These two requirements are in a trade-off relation
as a function of the thickness and 
the atomic number of the conversion material. 
We assume a gold foil with a thickness of 2 $\mu$m resulting in
a conversion efficiency of $1.3 \times 10^{-5}$ based on the
partially integrated cross section of the symmetric Bethe-Heitler process
as discussed in Eq.(\ref{eq18}).
Given the parameters in Tab.\ref{Tab1},
the expected number of conversion pairs per shot is 0.84,
which is close to unity.

\subsection{Capability to extract phase retardation}
The feasibility of extracting phase retardation with the detector 
configuration illustrated in Fig.\ref{Fig3} 
was evaluated using the Geant4 simulation toolkit\cite{geant1,geant2}.
In the following evaluation, for simplicity, we assume
a total number of conversion pairs as $10^{4}$ with a single pair
production per shot, which is likely achievable in
high-intensity laser facilities such as ELI~\cite{eli},
where 10~PW laser pulses are available with one shot per minute
resulting in 9 days in order to exceed $10^4$ pairs with
the expectation value of 0.84 pairs per shot.
A static magnetic field of 0.6 T over 12~cm was assumed 
just behind the converter which is enough to measure the sub-GeV momenta 
of the charge-separated electrons and positrons. 
To provide this field, a permanent-magnet-based dipole would be preferable from 
the point of view of the compactness and homogeneity of the field
in order to allow us to arbitrarily rotate the magnet system together 
with the set of sensors around the $z$-axis.
The three-layered position sensors made of silicon pixels are located 
downstream of the magnet system with the total length of the
polarimeter of 25~cm from the converter.
The pixel size and the thickness of the pixel sensor were assumed to 
be 20~$\mu$m and 50~$\mu$m, respectively.

As a result of the simulation, we found that 
the $\gamma$-ray energy can be reconstructed 
from the measured momenta of an $e^{-}e^{+}$ pair with
an energy resolution of 7.8$\%$ at $\omega = 1$~GeV. 
This energy resolution is sufficient to select a narrow enough energy range to 
guarantee the high degree of linear polarization of incident 
1.0~GeV $\gamma$-rays based on the offline selection of pairs,
because multi-photon absorption $n=2$ and $n=4$ cases give energies
0.726-0.730~GeV and 1.268-1.275~GeV in $\psi*<0.1$, respectively,
which can be discriminated from the $n=3$ case 1.015-1.021~GeV
with the 7.8\% resolution.

\begin{figure}
\includegraphics[scale=0.5]{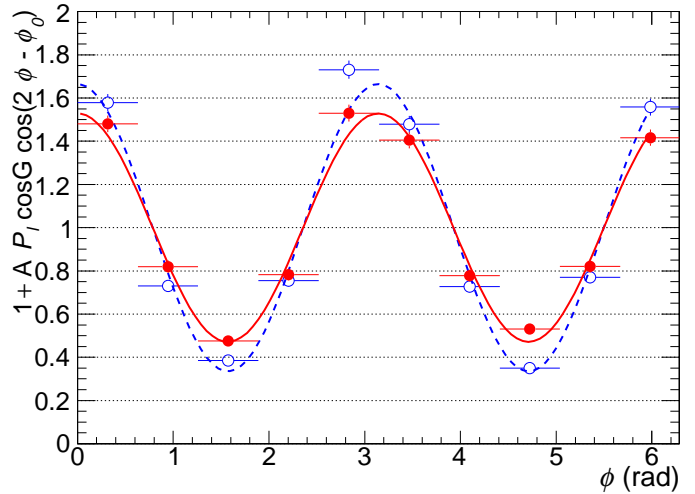}% Here is how to import EPS art
\caption{\label{Fig4} The angular distributions of $e^{-}e^{+}$ emission 
planes with respect the reference plane ($\phi_0 = 0$).
The open blue and closed red points depict the angular distributions
for $G=0$ and $0.72$, respectively,
with fitting results based on Eq.(\ref{eq_pol}).
The total number of $e^{-}e^{+}$ pairs was assumed to be $10^4$ 
with a single pair production per shot in this simulation.
The vertical error bars show statistical errors in the simulation.
}
\end{figure}

Figure \ref{Fig4} shows reconstructed angular distributions of
pair emission planes with respect to the reference plane ($\phi_0 = 0$).
We assume the creation of a single $e^{-}e^{+}$ pair per shot 
via the conversion process with a total pair statistics of 
$10^{4}$ in this simulation. 

The open blue and closed red points depict the angular distributions
for $G=0$ and $0.72$, respectively. 
The raw amplitudes of the angular distributions fit 
with Eq.(\ref{eq_pol}) are obtained as
$A(0.72)P_l\cos(0.72) = 0.529 \pm 0.013$ and
$A(0)P_l\cos(0) = 0.665 \pm 0.011$
where the errors include statistical errors and also
biases from the effect of the finite sensor segment and 
the track reconstruction algorithm.
The analyzing power $A$ has a monotonic $G$ dependence and 
we can evaluate them as A(0.72) = 0.725 and A(0) = 0.686 from 
the simulation in advance.

The reduction of the raw amplitude in $G=0.72$ 
from that of the null phase retardation case, 0.136,
can reach a high enough significance level
compared to the error size of the null retardation case, 0.011.
Therefore, we can declare the observation of phase retardation 
via vacuum birefringence given the statistics of $10^4$ pairs
by this method.

The averaged phase retardation $\langle G \rangle$ can then be extracted from
$\langle G \rangle = \cos^{-1}\left(P_l\cos(0.72)/P_l\cos(0)\right) = 0.720 \pm 0.034$ 
with the same statistics after correcting the analyzing power biases 
at the different $G$ values. 
We note that experiments do not necessarily have to quantify $P_l$ precisely 
because $P_l$ should be common to $G=0$ and $G=0.72$ cases and 
systematically canceled out.
From this simulation result, we evaluate that
the accuracy of the reconstructed $\langle G \rangle$ can reach 4.7\%.

\subsection{Possible sources of depolarization}
We consider several background sources which possibly change 
the degree of linear polarization of probe $\gamma$-rays 
before entering into the gold converter of the polarimeter 
with $d=1000$~cm in Fig.\ref{Fig1}.

The dominant background source would come from the mixture of two
linear polarization states in the nonlinear Compton scattering process.
How to correct the effect of $P_l$ of probe photons
has been already discussed in the previous subsection. 

The remaining contributions are
from possible interactions characterized by
individual cross sections $\sigma_{\gamma A}$
between $\gamma$-rays and residual atoms $A$
in the vacuum system along the distance $d$. 
The number of interacting $\gamma$-rays
is approximated as $N_{int} \sim \sigma_{\gamma A} n_{A} d$
with number density of residual atoms $n_A$.
A typical vacuum system maintained at $\sim 10^{-5}$~Pa results
in $n_A \sim 10^{10} \mbox{cm}^{-3}$ compared to
$n_A \sim 10^{20} \mbox{cm}^{-3}$ in the atmospheric pressure.
The possible interactions between GeV probe photons and atoms are
pair creations, Compton scattering, and  
Delbr\"{u}ck scattering~\cite{Delbrueck}.
The first two processes eventually absorb probe photons
or change the probe photon energy drastically, hence, 
they can be no serious background for the phase retardation measurement
as long as the narrow energy range and limited conversion points on
the converter of the polarimeter are imposed in the measurement. 
The interaction resulting in non-absorbed photons with the same energy 
as the generated energy at the Compton scattering vertex is thus limited to 
forward Delbr\"{u}ck scattering via $\gamma + A \rightarrow \gamma + A$. 

The differential forward Delbr\"{u}ck scattering cross section per solid angle
for high energy photons close to GeV is expected to be described as
$d\sigma_{\gamma A}/d\Omega = {\cal A}^2(\alpha Z)^4 (r_0)^2 
\sim {\cal A}^2(\alpha Z)^4 0.1$~b
with the classical electron radius $r_0$~\cite{FowawrdDelbrueck}.
The scattering amplitude is evaluated as ${\cal A} \sim 10^3$
at maximum for 1~GeV~\cite{FowawrdDelbrueck}. 
Even for Kr($Z=54$), corresponding highest $Z$ in the air, 
$d\sigma_{\gamma A}/d\Omega \sim (10^3)^2 (54/137)^4 0.1 \sim 10^3$~b at most.
For the assumed $d$ and $n_A$, we expect 
$N_{int} \sim 10^{-21} [\mbox{cm}^2] 10^{10} [\mbox{cm}^{-3}] 10^3 [\mbox{cm}]
= 10^{-8}$ per shot,
which is negligible with respect to $6.5 \times 10^4$ probe photons per shot
for the polarization measurement. 
Furthermore, by taking following facts into account:
i) the proper abundance of Kr as well as the same effects
from the other residual atoms with lower $Z$ in the air, 
ii) no reason to expect that residual atoms are polarized with respect to
the incident polarization plane of probe photons over the entire length $d$, 
and iii) very narrow solid angle in front of the polarimeter, we expect that
the depolarization effect by Delbr\"{u}ck scattering is totally negligible.

The robustness to the background contributions is 
one of advantages to use GeV probe photons compared to,
for example, the case of the PVLAS experiment where
careful controls of the vacuum pressure are required 
because eV photons more often interact with residual atoms and the atoms
could be weakly polarized by the external static magnetic field
along the entire path of probe photons.
We note, however, that tests of birefringence with different probe wavelengths
are essentially important in order to complete the measurement of the
dispersion relation in the laser-induced vacuum.

\section{Conclusion and prospects}
We have considered combining a 10~PW laser system
with 1~GeV linearly polarized probe $\gamma$-rays to enhance the sensitivity 
to the measurement of the laser-induced birefringence effect.
We have derived formulae to directly determine phase retardation
close to unity from pair-wise topology of the symmetric Bethe-Heitler process.
We conclude that if $10^4$ pairs are available, 
it is possible to observe the vacuum birefringence effect with the accuracy 
of 4.7\% for $\langle G \rangle=0.72$.
This result is based on a realistic set of laser parameters and 
potentially realizable statistics for 10~PW systems such as 
ELI projects~\cite{eli}.

Given the firm theoretical and experimental footing in the simplest QED case,
the proposed approach with the compact polarimeter design
would open up a new arena of fundamental physics
to explore more dynamical and complicated vacuum states 
realized in laboratories, astrophysical objects and 
possibly the early Universe. 

\begin{acknowledgments}
We are grateful to A. Ilderton for detailed discussions 
to quantify the flipping probabilities of $\gamma$-rays
and thank K. Seto and T. Moritaka for indepenent evalutaions on
the $\gamma$-ray yield.
We also thank S. Sakabe, M. Hashida, and S. Inoue for providing many insights into this subject.
The corresponding author,
K. Homma, acknowledges for the support by the Collaborative Research
Program of the Institute for Chemical Research,
Kyoto University (grants No. 2015-93, 2016-68, and 2017-67)
and the Grant-in-Aid for Scientific Research
no.15K13487, 16H01100, and 17H02897 from MEXT of Japan.
\end{acknowledgments}

\end{document}